\documentstyle[prl,aps,preprint]{revtex}
\begin{document}
\tightenlines
\draft
\title{Metallization of molecular hydrogen}
\author{Martin St\"adele and Richard M. Martin}
\address{Department of Physics, University of Illinois at Urbana-Champaign,
 Illinois 61801, USA}
\date{\today}
\maketitle

\begin{abstract}

We study metallization of molecular hydrogen under pressure
using exact exchange (EXX) Kohn-Sham density-functional theory in
order to avoid well-known underestimates of band gaps associated with standard
local-density or generalized-gradient approximations.
Compared with the standard methods, the EXX approach leads to considerably 
(1 - 2 eV) higher gaps and significant 
changes in the relative energies of different structures.
Metallization is predicted to occur at a density of $\stackrel{>}{\sim}$ 
0.6  mol/cm$^{3}$ (corresponding to a pressure of 
 $\stackrel{>}{\sim}$ 400 GPa), consistent with all previous measurements.


\end{abstract}
\narrowtext

Despite great efforts starting with the first theoretical predictions
in 1935~\cite{WIGNER}, the determination of the electronic and
structural properties of
hydrogen at high pressure
is still extremely
incomplete~\cite{HEMLEYREV}.
Experimentally, it is established that hydrogen transforms to
high pressure phases, but remains molecular up
to pressures of at least $\approx$ 200 GPa~\cite{191GPA216GPA}.
Metallization of solid hydrogen has been actively sought, but not yet
observed,
with one experimental team reporting no sign of metallization
up to 342 GPa~\cite{342GPA}.
It is widely assumed that metallization would occur either through a
structural transformation to an atomic
metallic phase, which involves dissociation of the H$_{2}$ molecules, or
through band overlap within the molecular phase itself.
This latter mechanism is supported by a recent experiment-based
equation of state~\cite{HCP} that, combined with
Quantum Monte Carlo (QMC) calculations for metallic atomic
 hydrogen~\cite{QMC}, yields an estimate for the dissociation pressure
of as much as 620 GPa~\cite{HCP}.

The theoretical situation is complicated by the fact that the
structures at high pressures are not known, together with
the well-known difficulties of quantitative predictions for
metal-insulator transitions. Various candidate structures for the 
high-pressure phases (called ``phase II'' or ``BSP'' below 
$\approx$ 150 GPa and 
 ``phase III'' or ``HA'' above $\approx$ 150 GPa)
have been proposed based on 
static~\cite{NAGARA,CMCA,KAXEU,NAGAO,XALPHA,MARTINS,STRAUS,IVO,ASH2000},
 and dynamic~\cite{KOH,ALAVI} density-functional calculations and on
QMC~\cite{QMC} investigations. Most of these structures have hexagonal
and orthorhombic unit cells with up to four molecules. However,
there are serious difficulties associated with the
estimates of metallization pressures.
The major problem is the well-known fact that the
local-density (LDA) or generalized gradient (GGA) approximations of
density-functional theory cause drastic underestimates of band gaps (by
typically 50 - 100 \%). This leads to much too low metallization pressures
and also affects the quality of LDA and GGA total energies that are
needed for the prediction of energetically favorable structures.
Previous work beyond the LDA and GGA was carried out within the
X$\alpha$ approximation~\cite{XALPHA}, the many-body GW framework in a
 first-principles~\cite{CHACHAM} and an approximate~\cite{ASH2000} 
formulation, and QMC simulations~\cite{QMC}. The
first two studies were limited to a simple hcp structure
with two molecules per cell oriented along the c-axis (called ``mhcp''
hereafter), which has been found to be energetically
unfavorable~\cite{NAGARA,CMCA,KAXEU,NAGAO};
the GW calculations\cite{ASH2000,CHACHAM} are not able to 
determine relative energies of structures.  
The QMC calculations
indicated qualitatively the problems with the LDA calculations but did not
determine gaps.

In this Letter, we present a first-principles
investigation of band-gap closure within
the molecular phase.  We employ the framework of exact exchange
density-functional theory (EXX), which has been shown recently to yield very
accurate band gaps {\em and} total energies for a large set of
semiconductors~\cite{EXX}.
The EXX method has crucial advantages for the present study.
Since it treats {\em exactly} all exchange-related quantities of
Kohn-Sham density-functional theory~\cite{KS},
it is inherently self-interaction free. This remedies
largely the band-gap underestimates that plague all LDA and GGA
calculations, without an artificial band-gap correction and in
a parameter-free way. Second, it yields band structures and total
energies {\em from the same calculation}, which we believe to be a key
requirement for the solution of the hydrogen problem.

In the EXX scheme, which is explained in detail in Ref. 
\onlinecite{EXX},  
 total energies $E_{tot}$ are obtained from the expression
\begin{equation} \label{EN}
E_{tot} = T_{0} + E_{el-prot} + E_{H} + E_{x} + E_{c}.
\end{equation}
Here $T_{0}$ is the noninteracting kinetic energy, $E_{el-prot}$ is the
interaction energy between the electrons and the protons, $E_{H}$ is the
Hartree energy, $E_{x}$ the exact exchange energy,
and $E_{c}$ denotes the correlation energy, which is the
only quantity that has to be approximated (the LDA is used in this
work). Band structures $\{\varepsilon_{n{\bf k}}\}$ and wavefunctions
$\phi_{n {\bf k}}$ for states with band index $n$ and wavevector ${\bf k}$
are obtained from the Kohn-Sham equations
\begin{equation} \label{POT}
\left( -\frac{\nabla^{2}}{2} + V_{prot}({\bf r}) + V_{H}({\bf r}) +
V_{x}({\bf r}) + V_{c}({\bf r}) \right)  \phi_{n {\bf k}} ({\bf r}) =
  \varepsilon_{n {\bf k}} \phi_{n {\bf k}}({\bf r}),
\end{equation}
where $V_{prot}({\bf r})$ is the potential due to the protons,
$V_{H}({\bf r})$ is the Hartree potential, and $V_{c}({\bf r}) =
\delta E_{c}/\delta n({\bf r})$ with the density $n({\bf r}) = 2
\sum_{n{\bf k}}^{occ} |\phi_{n{\bf k}}({\bf r})|^{2}$. The crucial 
part of the EXX scheme is the construction of the exact local 
exchange potential $V_{x}({\bf r}) = \delta E_{x}/\delta n({\bf 
r})$ as the functional derivative of the exact exchange energy with 
respect to the density~\cite{EXX}. Since this requires repeated 
computation of nonlocal-exchange integrals and a linear-response 
function, an EXX calculation is much more demanding than standard 
LDA or GGA methods. The present calculations were performed using 
the bare proton potential for hydrogen and plane-wave basis sets 
with a kinetic energy cutoff of 36 Ry that has been employed in 
previous calculations on H$_{2}$~\cite{XALPHA,MARTINS}. We have 
also performed  tests with cutoff energies of 60 Ry and observed 
that band gaps $\varepsilon_{gap}$ and total energy differences 
$\Delta E_{tot}$ among different structures were changed only by a 
few hundredths of an eV and a few tenths of a mRy/molecule, 
respectively. Dense ${\bf k}$-point meshes with $N_{\bf k} \approx 
3500/N_{at}$ special points in the Brillouin zone were employed 
($N_{at}$ denotes the number of protons in the unit cell). This 
guarantees convergence of $\Delta E_{tot}$ better than 1 
mRy/molecule.

First we show how the EXX band gaps compare with LDA and GW
band gaps over a wide range of densities, as depicted in
Fig.~\ref{FIGLIMIT} for the mhcp structure with the bond length
and c/a ratio fixed at the values determined from LDA and
 extrapolations of X-ray data~\cite{CHACHAM} (the only case for which GW
calculations have been done). We can recognize several salient features:
(i) EXX gaps are about 1.5 - 2 eV larger than the LDA gaps for 
all densities. Consequently, the EXX metallization density is 
considerably higher than for LDA. Similar corrections to band gaps 
(1 - 2 eV) have previously been reported for 
semiconductors~\cite{EXX,Kotani}. (ii) EXX and LDA gaps are almost
linear functions of density, which holds also for the gaps of the other
structures considered below.
(iii) a linear extrapolation of the EXX
data to zero density (isolated H$_{2}$ molecules) yields a gap of 11.4
eV, close to the weighted average of the lowest
experimental singlet and triplet excitation energies of the H$_{2}$
molecule~\cite{H2MOL}, 11.6 eV (indicated by the left cross in
Fig.~\ref{FIGLIMIT}).

The last point is in agreement with recent work on isolated 
noble-gas atoms~\cite{UMRIGARRESTA}: the differences between the 
highest occupied eigenvalue and the unoccupied EXX Kohn-Sham 
eigenvalues are very good approximations to excitation energies. 
This can be attributed to (i) the correct asymptotic $-1/r$ 
behavior of the exact exchange potential $V_{x}({\bf r})$ in 
Eq.~(\ref{POT}), which causes the EXX spectrum of the unoccupied 
states to be a Rydberg series (with  energies $\varepsilon_{1} < 
\varepsilon_{2} < \ldots < \varepsilon_{\infty} = 0$, see inset of 
Fig.~\ref{FIGLIMIT}) and (ii) $\varepsilon_{\infty} - 
\varepsilon_{HOMO}$ equaling the negative of the ionization energy 
$I$~\cite{PERDEW}. Indeed, we find $\varepsilon_{1} - 
\varepsilon_{HOMO}$ in EXX to agree very well with the lowest 
experimental excitation energies (crosses in Fig.~\ref{FIGLIMIT}), 
both for the isolated hydrogen molecule and the molecular solid at 
low density. Thus, the quasiparticle band gap $E_{gap}$, defined 
as the difference of the ionization energy and the electron 
affinity~\cite{PERDEW},  $E_{gap} = I_{H_{2}}-A_{H_{2}} \approx 
I_{H_{2}}$, differs from the lowest EXX gap by an exciton binding 
energy $\varepsilon_{\infty} - \varepsilon_{1}$~\cite{disc}.   At 
high densities excitonic effects are reduced, so that we expect 
the real quasiparticle gaps to deviate only slightly from the EXX 
gaps, just as has been demonstrated for semiconductors~\cite{EXX}. 

For densities greater than 0.35 mol/cm$^{3}$, we have also carried out
EXX calculations on other
structures with hexagonal and orthorhombic unit cells (see
Fig.~\ref{FIGSTR}) that have been proposed
previously~\cite{NAGARA,CMCA,KAXEU} as possible lowest-energy
structures on the basis of LDA and GGA total-energy calculations. 
Here, the H$_{2}$ molecules are tilted with respect to the z
 direction by an angle $\alpha \approx 55^{o}$ and the $c/a$ ratio 
is approximately 1.58 (at high pressures). In
the structures denoted by $Cmc2_{1}^{\delta}$, the centers of the
molecules are displaced from ideal hcp sites by a distance $\delta$ (we
normalize $\delta$ such that $Cmc2_{1}^{1}$ coincides with the $Cmca$
structure). Proton coordinates derived from LDA calculations for these
structures~\cite{NAGARA,IVO} were used as input for the present EXX
calculations since a complete unit-cell relaxation within the EXX scheme
is computationally too demanding at present.

Figure~\ref{FIGGAPS} depicts the fundamental EXX band gaps of the
 structures of Fig.~\ref{FIGSTR} for densities between $n_{1}$ = 0.35 and
$n_{2}$ = 0.60 mol/cm$^{3}$.
[The corresponding pressures can be specified by our theoretical 
calculations or by using an extrapolated experimental equation of 
state~\cite{HCP,Hemley,evans}. We find that at the densities $n_{1}$ 
and $n_{2}$, the theoretical 
pressures ($P_{1}$ and $P_{2}$) are close to those of Ref.~\cite{Hemley}, 
corresponding to $100$ GPa and $400$ GPa; 
Ref.~\cite{HCP} leads to much lower pressures at high 
density ($P_{1} = 100$ GPa,  $P_{2} = 325$ GPa), 
whereas Ref.~\cite{evans} gives 
higher pressures ($P_{1} = 115$ and $P_{2} = 500$ GPa).] 
For the $Cmc2_{1}^{0.5}$, 
$Pca2_{1}$, $Cmc2_{1}^{0}$, and $P2_{1}/c$ structures, we obtain 
metallization densities of 
0.468, 0.535, 0.537, and 0.542 mol/cm$^{3}$, respectively. Note 
that the use of LDA coordinates means at high pressure a 
bondlength of $r_{0} \approx$ 1.45 a.u. We have verified that 
using the experimental ($r_{0}^{Expt.} = 1.40$ a.u.) or EXX 
($r_{0}^{EXX} \approx 1.38$ a.u.) bondlength  of the isolated 
H$_{2}$ molecule~\cite{KIM} increases calculated band gaps by about 0.6 
and 0.9 eV, respectively. For the $P2_{1}/c$ structure, this is 
indicated 
 by the thin dashed lines in Fig.~\ref{FIGGAPS}.
The larger bondlength causes the predicted
 metallization density to increase up to 0.58 mol/cm$^{3}$,
corresponding to a calculated pressure of 375 GPa.

The EXX scheme predicts that the three structures with the largest gaps
($P2_{1}/c$, $Cmc2_{1}^{0}$, and $Pca2_{1}$) are the most stable ones.
Our results are reported in Fig.~\ref{FIGETOT} which
 shows the total energy differences
among the structures for densities up to 0.62 mol/cm$^{3}$. A key 
result of the utilization of the  EXX functional is that the 
metallic $Cmca$ structure becomes more stable than the insulating 
phases only above 
a density of 0.61 mol/cm$^{3}$ (calculated pressure  415 GPa). 
In contrast, LDA~\cite{ASH2000} and GGA calculations~\cite{KOH} 
find this to 
be the most stable structure at much 
lower density (at quoted pressures of P $\approx$ 140 and 180 GPa). 
Such  a low metallization pressure is in severe disagreement with 
experiment, and we believe the problem is a consequence of the 
erroneous LDA and GGA band gaps that indirectly affect the 
total energy. The 
energy is decreased by populating the conduction states, an effect 
that occurs in the EXX calculations only upon much higher 
compression. However, the rule  ``the lower the energy, the wider 
the band gap''~\cite{KAXEU} is not exactly obeyed: the most stable 
structure $Pca2_{1}$ has only the third  highest band gap. We find 
$Pca2_{1}$ to be more stable than $P2_{1}/c$ for all densities, in 
agreement with the LDA results of Ref.~\onlinecite{NAGAO}, but in 
disagreement with the LDA and GGA calculations of 
Refs.~\onlinecite{ASH2000} and~\onlinecite{KOH} that slightly 
favor $P2_{1}/c$.

Zero point motion of the protons is a very difficult problem that 
has been the subject of much debate.  For the present purposes 
there are three effects. First, the pressure is increased (by 
approximately 10\%~\cite{QMC}). Second, the band gaps may change.  
The three most stable structures have gaps that differ only by a 
few tenths of an eV. One might interpret this as an estimate of 
the influence of zero-point motion which is expected to average 
over low-energy structures. Tight-binding calculations on large 
cells with hydrogen molecules in disordered 
arrangements~\cite{CHACHAMTB} indicate effects that are similarly 
small. However, as the gaps become small, of order of the vibron 
energies $\approx$ 0.5 eV, 
 we expect the zero point motion to
increase the gaps by a dynamic level-repulsion effect. Third, 
relative energies of different structures are changed. QMC 
calculations~\cite{QMC} and work by Straus and 
Ashcroft~\cite{STRAUS} suggest that zero-point motion favors 
isotropic structures (in our case, $Pca2_{1}$, $P2_{1}/c$, and 
$Cmc2_{1}^{0}$) with respect to anisotropic ones like $Cmca$. 
Including all these effects, we expect the metallization pressure
 to increase to $\stackrel{>}{\sim}$ 400 GPa.

Another possibility is the metallization by a structural transition to a
possible monatomic phase.  A comparison of
enthalpies derived from various experimental equations of state with
QMC calculations~\cite{QMC} for
hydrogen in the diamond phase yields dissociation pressures between 300
and 620 GPa~\cite{HCP,QMC}.
 The large range is due to the extreme sensitivity
of the transition point to the form of the equation of state. Thus 
we can only conclude that our calculated metallization pressure is 
in the same range as possible transitions to other structures. However, 
this does not affect our main point that up to pressures of at 
least 400 GPa, molecular 
hydrogen is predicted to be stable and insulating.

In summary, we have investigated band gaps and total energies of
possible candidate structures for compressed molecular hydrogen using
a Kohn-Sham density-functional scheme (EXX) that treats exchange
interactions exactly. EXX leads to band gaps that are 1 - 2 eV higher
than in LDA 
(similar to gaps found recently using an approximate GW
approach~\cite{ASH2000}) and, in addition, predicts 
changes of the relative energies of structures near the
metal-insulator transition. 
In contrast to LDA and GGA calculations,
the energetically preferred structure has $Pca2_{1}$ symmetry
up to density 0.61 mol/cm$^3$ (pressure  $\approx$ 400 - 450 GPa).
In this structure there is  the possibility of metallization via 
band overlap, which is here found to occur at $\approx$ 400 GPa.
Above this pressure there are three 
possibilities: a metallic molecular phase as described here; some 
new molecular phase that is more stable and insulating; or a 
transition to an atomic phase expected to be metallic.

We acknowledge interesting discussions with W. Evans, A. G\"{o}rling, 
R. J. Hemley, J. Kohanoff, J. B. Krieger, P. Loubeyre, J. P. Perdew,
 I. F. Silvera,
I. Souza, and B. Tuttle.
 This work has been supported by the Office of Naval Research
under Grant No. N00014-98-1-0604.

\begin{figure}
\caption{\label{FIGLIMIT} Fundamental band gaps of the mhcp structure,
 calculated by the EXX, GW~\protect\onlinecite{CHACHAM}, and LDA methods. 
The squares represent
experimental estimates for the band gap~\protect\onlinecite{EXPERIMENT},
the crosses denote lowest experimental excitation 
energies~\protect\onlinecite{H2MOL,EXPERIMENT}.
The inset shows qualitatively the EXX eigenvalue
spectrum in the zero-density limit.}
\end{figure}

\begin{figure}
\caption{Possible ground-state structures for solid H$_{2}$
at high pressures, projected onto the $xy$ plane. Full (empty) arrows
represent molecules centered on the $c$ ($c$/2) plane and pointing
towards the positive-z hemisphere. Though some of the structures have a
two-molecule minimum basis, we have indicated the rectangular cross
sections of orthorhombic four-molecule unit cells for the ease of
comparison.  \label{FIGSTR}}
\end{figure}

\begin{figure}
\caption{Fundamental EXX band gaps
of various candidate structures for molecular hydrogen
as a function of density.
The thin long-dashed (short-dashed) line indicates the gaps of the
$P2_{1}/c$ structure obtained using the EXX
(experimental) bondlength of the isolated molecule. \label{FIGGAPS}}
\end{figure}

\begin{figure}
\caption{EXX total energies of possible structures of molecular hydrogen,
relative to the energy of the $Pca2_{1}$
structure.  \label{FIGETOT}}
\end{figure}

\end{document}